\documentclass[runningheads]{llncs}
\usepackage{graphicx}
\usepackage{amsmath}
\usepackage{amssymb}
\begin{document}
\title{Segmentation Loss Odyssey}
%
%
\author{Jun Ma}
\authorrunning{J. Ma et al.}
\institute{Department of Mathematics, Nanjing University of Science and Technology}
%
%
\maketitle              
\begin{abstract}
Loss functions are one of the crucial ingredients in deep learning-based medical image segmentation methods. Many loss functions have been proposed in existing literature, but are studied separately or only investigated with few other losses.
In this paper, we present a systematic taxonomy to sort existing loss functions into four meaningful categories. This helps to reveal links and fundamental similarities between them. Moreover, we explore the relationship between the traditional region-based and the more recent boundary-based loss functions.
The PyTorch implementations of these loss functions are publicly available at \url{https://github.com/JunMa11/SegLoss}.

\keywords{Loss function  \and Segmentation \and Deep learning \and Taxonomy.}
\end{abstract}
\section{Background and Notation}
\begin{figure}
\center
\includegraphics[scale=0.4]{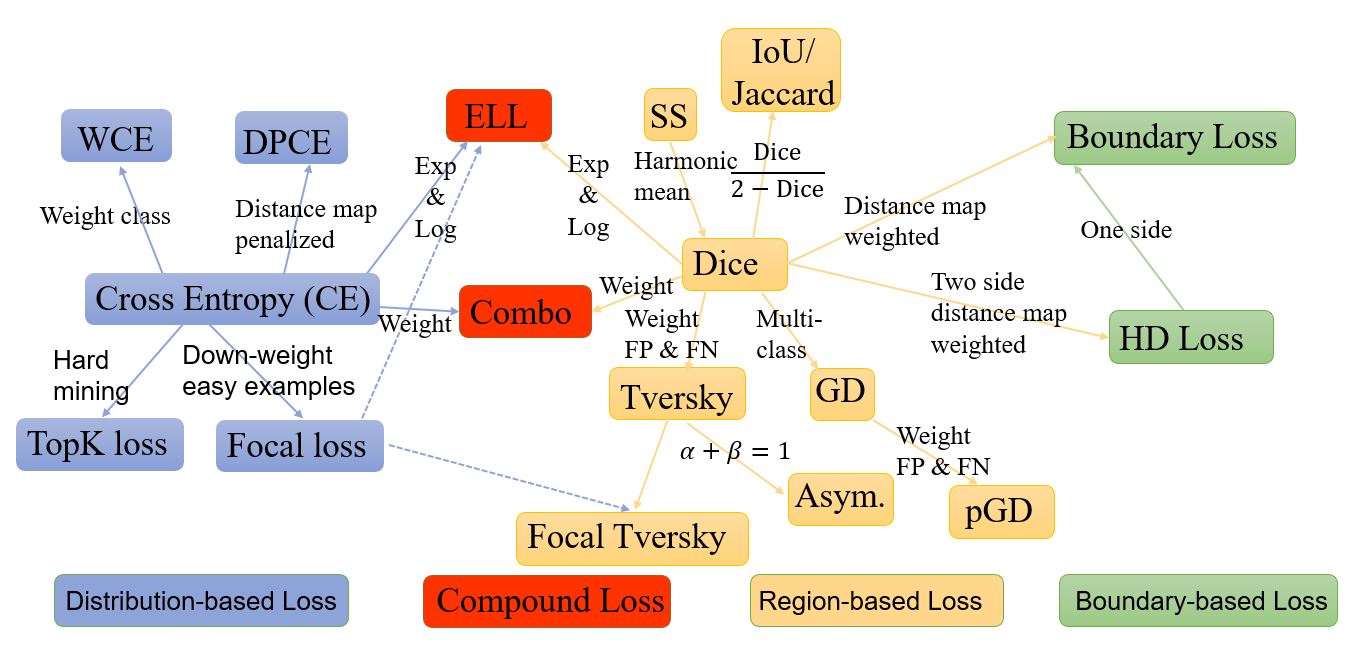}
\caption{Overview and relationship among the existing loss functions} \label{fig1}
\end{figure}

The role of a loss function is to evaluate how well the predicted segmentation matches the ground truth, which is an important part in a deep learning pipeline. Over the past five years, various loss functions have been proposed for medical image segmentation.
In this paper, we present an overview (Fig. \ref{fig1}) of how the current loss functions can be sorted into meaningful categories according to their optimization objective. We distinguish between loss functions that minimize the mismatch in distribution, region, boundary  or some combination of these.
Moreover, we will  explore the connections between these loss functions.

In general, the loss function $L$ takes the form of expected risk:
\begin{equation}
    L = E(G, S)
\end{equation}
where $G=\{g_i\}$ and $S=\{s_i\}$ are ground truth and predicted segmentation, respectively.
Let $\Omega$ denotes image domain with $N$ pixels and $C$ classes.

\section{Distribution-based Loss}
\label{DistributionLoss}
Distribution-based loss functions aim to minimize dissimilarity between  two  distributions. The most fundamental function in this category is cross entropy; all other functions are derived from cross entropy.

\subsection{Cross entropy }
Cross entropy (CE) is derived from Kullback-Leibler (KL) divergence, which is a measure of dissimilarity between two distributions. For common machine learning tasks, the data distribution is given by the training set. Thus, minimizing KL divergence is equivalent to minimizing CE. Cross entropy is difined by

\begin{equation}
    L_{CE} = -\frac{1}{N}\sum_{i=1}^{N}\sum_{c=1}^{C}g_{i}^{c}logs_{i}^{c}
\end{equation}
where $g_i^c$ is binary indicator if class label
$c$ is the correct classification for pixel $i$, and $s_i^c$ is the corresponding predicted probability.



Weighted cross entropy (WCE) \cite{wce2015} is a commonly used CE extension,

\begin{equation}
    L_{WCE} = -\frac{1}{N}\sum_{i=1}^{N}\sum_{c=1}^{C}w_cg_{i}^{c}logs_{i}^{c}
\end{equation}
where $w_c$ is the weight for each class. Usually, $w_c$ is inversely proportional to the class frequencies which can penalize majority classes.

\subsection{TopK loss}
TopK loss \cite{wu2016topK} aims to force networks to focus on hard samples during training.

\begin{equation}
    L_{TopK} = - \frac{1}{\sum_{i=1}^{N}\sum_{c=1}^{C} 1\{g_i = c \; and \; p_{ic} < t\}} (\sum_{i=1}^{N}\sum_{c=1}^{C}1 \{y_i = c \; and \; s_{i}^{c}<t\} logs_{i}^{c})
\end{equation}
where $t\in (0,1]$ is a threshold and $1\{\dots\}$ is the binary indicator function. In other words, ``easy'' pixels, i.e., the pixels with losses below $t$, are dropped as they are easily classified by the current model.

\subsection{Focal loss }
Focal loss \cite{focal2017} adapts the standard CE to deal with extreme foreground-background class imbalance, where the loss assigned to well-classified examples is reduced.

\begin{equation}
    L_{FL} = - \frac{1}{N}\sum_{i=1}^{N}\sum_{c}^{C} (1 - s_i^c)^{\gamma} g_{i}^{c}logs_i^c
\end{equation}

\subsection{Distance map penalized cross entropy loss (DPCE)}
DPCE loss \cite{MIDL19DisWeight} weights cross entropy by distance maps which are derived from ground truth masks. It aims to guide the network's focus towards hard-to-segment boundary regions.

\begin{equation}
    L_{DPCE} = -\frac{1}{N}(1+D)\circ \sum_{i=1}^{N}\sum_{c=1}^{C} g_i^c log s_i^c
\end{equation}
where $D$ is the distance penalty term, and $\circ$ is the Hadamard product. Specifically, $D$ is generated by computing distance transform of ground truth and then reverting them.

\section{Region-based Loss}
Region-based loss functions aim to minimize the mismatch or maximize the overlap regions between ground truth $G$ and predicted segmentation $S$. The key element is the Dice loss.

\subsection{Sensitivity-specificity loss}
Sensitivity-specificity loss \cite{ss2015} addresses imbalanced problems by weighting the specificity higher.
\begin{equation}
    L_{SS} = w\frac{\sum_{i=1}^{N}\sum_{c}^{C}(g_i^c - s_i^c)^{2}g_{i}^c}{\sum_{i=1}^{N}\sum_{c}^{C}g_i^c + \epsilon} + (1-w)\frac{\sum_{i=1}^{N}\sum_{c}^{C}(g_i^c - s_i^c)^{2}(1-g_{i}^c)}{\sum_{i=1}^{N}\sum_{c=1}^{C}(1-g_i^c) + \epsilon}
\end{equation}
where the parameter $w$ controls the trade-off between sensitivity (the first term) and sensitivity (the second term).


\subsection{Dice loss}
Dice loss \cite{dice2016} can directly optimize the Dice coefficient which is the most commonly used segmentation evaluation metric. Unlike weighted cross entropy, it does not require class re-weighting for imbalanced segmentation tasks.

\begin{equation}
    L_{Dice} = 1- \frac{2\sum_{i=1}^{N}\sum_{c=1}^{C}g_{i}^{c}s_{i}^{c}}{\sum_{i=1}^{N}\sum_{c=1}^{C}g_{i}^{c2} + \sum_{i=1}^{N}\sum_{c=1}^{C}s_i^{c2}}
\end{equation}

\subsection{IoU loss}
IoU loss \cite{rahman2016optimizing}, similar to Dice loss, is also used to directly optimize the object category segmentation metric.

\begin{equation}
    L_{IoU} = 1 - \frac{\sum_{i=1}^{N}\sum_{c=1}^{C} g_i^c  s_i^c}{\sum_{i=1}^{N}\sum_{c=1}^{C} (g_i^c + s_i^c - g_i^c s_i^c)}
\end{equation}

\subsection{Tversky loss }
To achieve a better trade-off between precision and recall, Tversky loss \cite{salehi2017tversky} reshapes Dice loss and emphasizes false negatives.

\begin{equation}\label{tversky}
    \begin{aligned}
        L_{Tversky} &= T(\alpha, \beta)\\
        &= \frac{\sum_{i=1}^{N}\sum_{c}^{C} g_i^c  s_i^c}{\sum_{i=1}^{N}\sum_{c}^{C} g_i^c  s_i^c + \alpha\sum_{i=1}^{N}\sum_{c}^{C} (1-g_i^c)  s_i^c + \beta\sum_{i=1}^{N}\sum_{c}^{C} g_i^c  (1-s_i^c)}
    \end{aligned}
\end{equation}
where $\alpha$ and $\beta$ are hyper-parameters which control the balance between false negatives and false positives.


\subsection{Generalized Dice loss }
Generalized Dice loss \cite{sudre2017generalised} is the multi-class extension of Dice loss where the weight of each class is inversely proportional to the label frequencies.

\begin{equation}
    L_{GD} = 1 - 2\frac{\sum_{c=1}^C w_c \sum_{i=1}^N g_{i}^c  s_{i}^c}{\sum_{c=1}^C w_c \sum_{i=1}^N (g_{i}^c + s_{i}^c)}
\end{equation}
where $w_c=\frac{1}{(\sum_{i=1}^N g_{i}^c)^2}$ is used to provide invariance to different label set properties.


\subsection{Focal Tversky loss }
Focal Tversky loss \cite{abraham2018novel} applies the concept of focal loss to focus on hard cases with low probabilities.

\begin{equation}
    L_{FTL} = (1 - L_{Tversky})^{\frac{1}{\gamma}}
\end{equation}
where $\gamma$ varies in the range $[1, 3]$.

\subsection{Asymmetric similarity loss }
Asymmetric similarity loss \cite{Asymmetric} introduces a weighting parameter $\beta$ to better adjust the weights of false positives and false negatives. It is also a special case of Tversky loss when $\alpha + \beta = 1$ in Eq. (\ref{tversky}).


\begin{equation}
    L_{Asym} = \frac{\sum_{i=1}^N g_{i}s_i}{\sum_{i=1}^N g_{i}s_i + \frac{\beta^2}{1+\beta^2}\sum_{i=1}^N g_i (1-s_i) + \frac{1}{1+\beta^2}\sum_{i=1}^N (1-g_i)s_i}
\end{equation}


\subsection{Penalty loss }
Penalty loss \cite{MIDL19Penalty} penalizes false negatives and false positives in generalized Dice loss $L_{GD}$.

\begin{equation}
  L_{pGD} = \frac{L_{GD}}{1+k(1-L_{GD})}
\end{equation}
where $k$ is a non-negative penalty coefficient. When $k=0$, $L_{pGD}$ is equivalent to generalized Dice loss. When $k>0$, $L_{pGD}$ gives additional weights to false positives and false negatives.

\section{Boundary-based Loss}
Boundary-based loss, a new type of loss function, aims to minimize the distance between ground truth and predicted segmentation.

\subsection{Boundary (BD) loss}
To compute the distance $Dist(\partial G, \partial S)$ between two boundaries in a differentiable way, boundary loss uses integrals over the boundary instead of unbalanced integrals over regions to mitigate the difficulties of highly unbalanced segmentation.
\begin{equation*}
\label{eq:distance}
\begin{split}
    Dist(\partial G, \partial S)
    &= \int _{\partial G} ||q_{\partial S}(p) - p||^2 dp\\
    &\approx2\int_{\Delta S}D_{G}(p) dp\\
    &=2(\int_{\Omega}\phi_{G}(p)s(p)dp - \int_{\Omega}\phi_{G}(p) g(p)dp))
\end{split}
\end{equation*}
where $\Delta M=(S/G)\cup(G/S)$ is the mismatch part between ground truth $G$ and segmentation $S$, $D_G(p)$ is the distance map of ground truth, $s(p)$ and $g(p)$ are binary indicator function.
$\phi_G$ is the level set representation of boundary: $\phi_G = -D_G(q)$ if $q\in G$, and $\phi_G = D_G(q)$ otherwise.
Then $s(p)$ is replaced by the network softmax probability outputs $s_\theta (p)$ to form a trainable function.
The last term is omitted as it is independent to the network parameters.
Finally, we obtain the following boundary loss function:
\begin{equation*}
    L_{BD} = \int_{\Omega}\phi_{G}(p)s_{\theta}(p)dp
\end{equation*}

\subsection{Hausdorff Distance (HD) loss}
Since minimizing HD directly is intractable and could lead to unstable training,
Karimi et al. \cite{karimi2019HDLoss} show that it can be approximated by the distance transforms of ground truth and predicted segmentation. Further more, the network can be trained with following HD loss function\footnote{HD loss is an estimation of Hausdorff distance, not exact Hausdorff distance.} to reduce HD:
\begin{equation*}
    L_{HD_{DT}} = \frac{1}{N} \sum_{i=1}^{N} [(s_i - g_i) \cdot (d_{Gi}^2 + d_{Si}^2)]
\end{equation*}
where $d_G$ and $d_S$ are distance transforms of ground truth and segmentation.

\subsection{Connection among Dice loss, BD loss and HD loss}
For ease of illustration, we focus on binary segmentation. The three loss functions can be rewrited as following formulations:

Dice loss: $L_{Dice} = 1- \frac{2|G\cap S|}{|G|+|S|} =\frac{\Delta M}{|G|+|S|} = \frac{\Delta M}{\sum_{i=1}^{N}s_i + \sum_{i=1}^{N} g_i}$ \\

BD loss: $L_{BD}\approx \int_{\Delta M}D_{G}(p) dp$ \\

HD loss: $L_{HD-DT}\approx \frac{1}{|\Omega|}\sum_{\Omega} \Delta M \circ (D_G +  D_S)$\\

It can be found that all three loss functions aim to minimize the mismatch regions $\Delta M$ between ground truth and segmentation. The key difference among them is the weighting methods.
For Dice loss, the segmentation mismatch is weighted by the sum of the number of foreground pixels in the segmentation and the number of pixels in ground truth. In BD loss, it is weighted by the distance transform map of ground truth. HD loss not only uses the distance transform map of the ground truth for weighting, but also uses the distance transform map of the segmentation.

\section{Compound Loss}
\subsection{Combo loss}
Combo loss (\cite{isensee2019nnu}\cite{Combo18}) is the weighted sum between weighted CE and Dice loss.

\begin{equation}
\begin{split}
        L_{Conbo} &= \alpha (-\frac{1}{N} \sum_{i=1}^{N} \beta (g_i log s_i) + (1-\beta)[(1-g_i)log(1-s_i)]) \\
        &- (1-\alpha) (\frac{2\sum_{i=1}^{N}s_{i}g_{i} + \epsilon}{\sum_{i=1}^{N}s_i + \sum_{i=1}^{N}g_i + \epsilon})
\end{split}
\end{equation}

\subsection{Exponential Logarithmic loss (ELL)}
Wong et al. \cite{wong2018els} propose to make exponential and logarithmic transforms to both Dice loss an cross entropy loss. In this way, the network can be forced to intrinsically focus more on less accurately predicted structures.

\begin{equation}
    L_{ELL} = w_{Dice}E[(-log(Dice_c))^{\gamma Dice}] + w_{CE}E[w_{c}(-log(s_{i}^c))^{\gamma CE}]
\end{equation}
where $Dice_c = \frac{2\sum_{i=1}^N g_{i}^c  s_{i}^c + \epsilon}{\sum_{i=1}^N (g_{i}^c + s_{i}^c) + \epsilon}$.

\section{Relationships, Recommendations and Conclusions}
\textbf{Relationship among the loss functions:}
There are strong links among the existing loss functions as shown in Fig. \ref{fig1}. Most of the distribution-based and region-based loss functions are variants of cross entropy and Dice loss, respectively. Boundary-based losses are motivated by minimizing the distance between two boundaries, but we show some similarities to Dice loss, as both are computed with region-based methods. Compound loss is the combination of different types of loss functions.

\noindent \textbf{Recommendations for choosing loss functions:}
To the best of our knowledge, there has not yet been a comprehensive empirical comparison of all mentioned loss functions. Thus, it is hard to identify the best loss function, however, we can obtain some insights from existing literature.
Mildly imbalanced problems are well handled by Dice loss or generalized Dice loss \cite{sudre2017generalised}. Highly imbalanced segmentation tasks are much more difficult and require more robust loss functions.
Wong et al. \cite{wong2018els} show that exponential logarithmic loss function can achieve better performance than Dice loss and cross entropy on a brain segmentation task with 20 labels. Moreover, both Taghanak et al. \cite{Combo18} and Isensee et al. \cite{isensee2019nnu} highlight that the sum of Dice loss and cross entropy performs better than using them separately in many segmentation tasks. Recently, Kervadec et al. \cite{kervadec2018boundary} demonstrate that combining boundary loss with generalized Dice loss can address highly imbalanced segmentation tasks. Thus, for the \textbf{QUESTION:} which loss function should we choose for medical image segmentation tasks? The \textbf{ANSWER} is that, overall, using compound loss functions is a better choice. \\
\noindent \textbf{Conclusions:}
Our loss function taxonomy provides an overview of the existing loss functions and gives users a clear picture to assist in understanding the relationships among them.\\ 
\textbf{Future work} We will evaluate all the loss functions and present a loss function benchmark for medical image segmentation.

%
%
%
\bibliographystyle{splncs04}
\bibliography{lossfunctions}

\end{document}